\documentclass[hidelinks,10pt,twocolumn]{article}
\usepackage{hotnets-style}

\usepackage[pdftex]{graphicx}
\usepackage{booktabs}  % for tables
\usepackage{dcolumn}   % for tables
\usepackage{multirow}  % for tables
\usepackage{rotating}
\usepackage{xspace}
\usepackage{hyphenat}  % supplies \hyp{}, which tells tex that it can 
\usepackage[dvipsnames]{xcolor}
\usepackage{enumerate}
\usepackage{comment}

\usepackage[square,comma,numbers,sort&compress]{natbib}

\usepackage{wrapfig}
\usepackage{textcomp}
\usepackage{lastpage}
\usepackage{tabularx}
\usepackage{pifont}

\usepackage{wrapfig}

\usepackage{ulem}
\usepackage{colortbl} % colorful columns and rows
\usepackage{array}    % needed for 'b' argument in tabular preamble

\usepackage{dblfloatfix}

\usepackage{algpseudocode}
\algrenewcomment[1]{\hfill// #1}%
\algnotext{EndFunction}
\algnotext{EndFor}
\algnotext{EndIf}

\usepackage{amsmath,amscd}
\usepackage{amssymb}
\usepackage{amsfonts}
\usepackage{amsthm}

\algrenewcommand\algorithmicindent{1em}
\algrenewcommand{\algorithmiccomment}[1]{{\color{CadetBlue}\hfill// #1}}

\usepackage[noeka]{mathrmletter}

\newif\ifextended
\newif\iflongbatching
\newif\ifsubmission
\newif\ifelementary

\ifx\buildextended\undefined
\else
    \extendedtrue
\fi

\ifx\noeditingmarks\undefined
   \newcommand{\pgwrapper}[3]{\begingroup \color{#1} #2: #3 \endgroup}
   \newcommand{\pgwrapperb}[1]{\textbf{#1}}
\else
   \newcommand{\pgwrapperb}[1]{}
   \newcommand{\pgwrapper}[3]{}
\fi

\def\hn{\usefont{OT1}{phv}{mc}{n}\selectfont}

\newcommand{\mpfont}{\hn\scriptsize}

\ifx\noeditingmarks\undefined
    \newcommand{\MPworker}[2]{{\color{#1}\vrule\vrule}{\marginpar{\color{#1}\mpfont #2}}}
\else
    \newcommand{\MPworker}[2]{}
\fi

\ifx\noeditingmarks\undefined
    
\else
    
\fi

\newcommand{\cnb}{Carbonari and Beschastnikh\xspace}
\newcommand{\rackmmu}{Rack MMU\xspace}

\theoremstyle{definition}

\makeatletter
\renewcommand*{\@fnsymbol}[1]{\ensuremath{\ifcase#1\or \star\or \dagger\or \ddagger\or
   \mathsection\or \mathparagraph\or \|\or **\or \dagger\dagger
   \or \ddagger\ddagger \else\@ctrerr\fi}}
\makeatother

\makeatletter
\def\imod#1{\allowbreak\mkern10mu({\operator@font mod}\,\,#1)}
\makeatother


\def\compactify{\itemsep=0in \topsep=2pt \parsep=0.00in \partopsep=0pt
\leftmargin=2em}
\let\latexusecounter=\usecounter

\newenvironment{myitemize}%
  {\begin{list}{\labelitemi}{\itemsep3pt \topsep3pt \parsep0.00in
  \partopsep=3pt \leftmargin1.2em}}%
  {\end{list}}
  {\begin{list}{\labelitemi}{\itemsep1pt \topsep2pt \parsep0.00in
  \partopsep=0pt \leftmargin1.2em}}%
  {\end{list}}
  {\begin{list}{\labelitemi}{\itemsep2pt \topsep2pt \parsep0.00in
  \partopsep=0pt \leftmargin1.2em}}%
  {\end{list}}
  {\begin{list}{\threequartdash}{\itemsep3pt \topsep3pt \parsep0.00in
  \partopsep=3pt \leftmargin1.5em}}%
  {\end{list}}

\def\compactsortof{\itemsep=0in \topsep=2pt \parsep=0.00in \partopsep=0pt
\leftmargin=1.7em}

\def\compactsqueeze{\itemsep=0pt \topsep0pt \parsep=0ex \partopsep=0pt
\leftmargin=1.63em}

\ifx\normalpar\undefined
  
\else
  
\fi

\def\discretionaryslash{\discretionary{/}{}{/}}
{\catcode`\/\active
\gdef\URLprepare{\catcode`\/\active\let/\discretionaryslash
        \def~{\char`\~}}}%
\def\URL{\bgroup\URLprepare\realURL}%
\def\realURL#1{\tt #1\egroup}%

\begin{document}

\title{Disaggregation and the Application}
\author{
  Sebastian Angel \\ {\it \normalsize University of Pennsylvania} \and 
  Mihir Nanavati \\ {\it \normalsize Microsoft Research} \and 
  Siddhartha Sen \\ {\it \normalsize Microsoft Research}
}
\date{}

\maketitle

\begin{abstract}

This paper examines disaggregated data center architectures from the
  perspective of the applications that would run on these data
  centers, and challenges the abstractions that have been proposed
  to date.
In particular, we argue that operating systems for disaggregated data
  centers should not abstract disaggregated hardware resources, such
  as memory, compute, and storage away from applications, but should
  instead give them information about, and control over, these
  resources.
To this end, we propose additional OS abstractions and interfaces for
  disaggregation and show how they can improve data transfer in data
  parallel frameworks and speed up failure recovery in replicated,
  fault-tolerant applications.
This paper studies the technical challenges in providing applications
  with this additional functionality and advances several
  preliminary proposals to overcome these challenges.
\end{abstract}

\section{Introduction}\label{s:intro}

Disaggregation splits existing monolithic servers into a number of
  consolidated single-resource pools that communicate over a fast
  interconnect~\cite{kyathsandra13intel, asanovic14firebox,
  katrinis16rack, lim09disaggregated, lim11disaggregated,
  themachine15, yizhou18legoos}.
This model decouples individual hardware resources, including
  tightly bound ones such as processors and memory, and enables the
  creation of ``logical servers'' with atypical hardware
  configurations.
Disaggregation has long been the norm for disk-based
  storage~\cite{katz92high} because it allows individual 
  resources to scale, evolve, and be managed independently of one
  another.
In this paper, we target the new trend of memory disaggregation.

Existing works on disaggregated data centers (DDCs) have focused
  primarily on the \emph{operational benefits} of disaggregation---it allows
  resources to be packed more densely and improves utilization by
  eliminating the bin-packing problem.
As a result, these works strive to preserve existing abstractions and 
  interfaces and propose runtimes and OSes that make the unique 
  characteristics of DDCs transparent to 
  applications~\cite{carbonari17tolerating, yizhou18legoos}. 
The implicit underlying assumption in these works is that from the perspective 
  of the OS, the distributed nature of processors and memory is an 
  inconvenient truth of the underlying hardware, much like paging or 
  interrupts, that should be abstracted away from applications.

Our position is that the disaggregated nature of DDCs is not just a
  hardware trend to be tolerated and 
  abstracted away to support legacy
  applications, but rather one that should be \textit{exposed to applications 
  and exploited for their benefit}.
We draw inspiration from decades-old distributed shared memory
  systems (which conceptually closely resemble disaggregation)
  where early attempts at full transparency quickly gave way to weaker
  consistency and more restrictive programming models for performance
  reasons~\cite{hudaklidsm, treadmarks, opal, cashmere}.
While the driving rationale for externalizing memory has changed,
  along with the underlying hardware and target applications, we believe
  that co-designing applications and disaggregated operating systems
  remains an attractive proposition.

Two properties of disaggregated hardware with 
  potential to benefit applications are the ability to \textit{reassign
  memory} by dynamically reconfiguring the mapping between processors
  and memory and the \textit{failure independence} of different hardware
  components (i.e., the fact that processors may fail without the
  associated memory failing or vice versa).
Memory reassignment can be leveraged by applications performing
  bulk data transfers across the network to achieve zero-copy operations
  by remapping memory from the source to the destination, or during
  processor failures to find orphaned memory a new home.
Failure independence also allows processors to be useful despite
  memory failures by acting as fast and reliable failure 
  informers~\cite{aguilera09no} and triggering recovery protocols.

We target data center applications that are logically cohesive, but
  physically distributed across multiple co-operating
  instances---examples of these include most microservice-based 
  applications, data parallel frameworks,
  distributed data stores, and fault-tolerant locking and metadata
  services---and propose extending existing OSes for disaggregated
  systems, such as LegoOS~\cite{yizhou18legoos}, with primitives
  for memory reassignment and failure notification.
Below is a discussion of the proposed primitive operations and
  the challenges in implementing them, all of which are exacerbated by
  the fact that the exact nature of disaggregation and the functionality
  of each component is in flux (\S\ref{s:bg}).

\begin{myitemize}
  \item \emph{Memory grant.} This is a voluntary memory
      reassignment called by a source application instance to yield its
      memory pages and move them to a destination application instance.
      This reassignment requires a degree of flexibility from the
      interconnect, which must be able to handle modifying memory
      mappings quickly and at fine granularities.

  \item \emph{Memory steal.} This is an involuntary reassignment of
      memory from one application instance to another. While similar
      to a memory grant from the perspective of the interconnect, a
      key difference is that the source application instance may not have
      any prior warning. Since volatile state can now transcend an application 
      instance, the programming model needs to guarantee 
      \textit{crash consistency} to ensure that state is semantically coherent 
      at all times.

  \item \emph{Failure notification.} An application instance can opt to
      receive notifications for memory failures or it can register other 
      instances to automatically be notified in such cases. This
      requires making failure information visible to applications,
      as well as retaining group membership at the 
      processor so other instances can be notified if the
      local instance cannot handle or mask the memory failure. 
\end{myitemize}

Data parallel frameworks, such as MapReduce and Dryad, can use
  these primitives to eliminate unnecessary data transfer during shuffles or
  between nodes in the data flow graph, while
  Chubby~\cite{burrows06chubby} and other applications
  based on Paxos~\cite{lamport98part} can recover and reassign the
  committed state machine from a failed replica.
In addition, early detection of memory failures can trigger recovery
  mechanisms without waiting for conservative end-to-end timeouts.
While this paper focuses on these two applications, we
believe that the interfaces are broad enough to benefit other applications.
For example, scalable data stores, such as Redis or memcached,
  could use memory grants to delegate part of their key space to new
  instances sans copying, while microservice-based applications can
  use grants and steals to achieve performance comparable to monolithic 
  services and still retain some modularity. 

This paper is largely speculative and poses more questions than it
  answers.
We discuss some operating system primitives that empower
  forward-looking applications to benefit from disaggregation (while
  allowing legacy applications to remain oblivious to it) and describe
  how these primitives could be implemented and used in the context of an
  abstract disaggregated data center that resembles existing designs.
Our hope is to foster a broader discussion around disaggregation, not
  from the perspective of operators, but as an opportunity, and also a
  challenge, for systems and application developers.

\section{DDC architecture and resource allocation}\label{s:bg}

\begin{figure}
\centering
\includegraphics[width=\columnwidth]{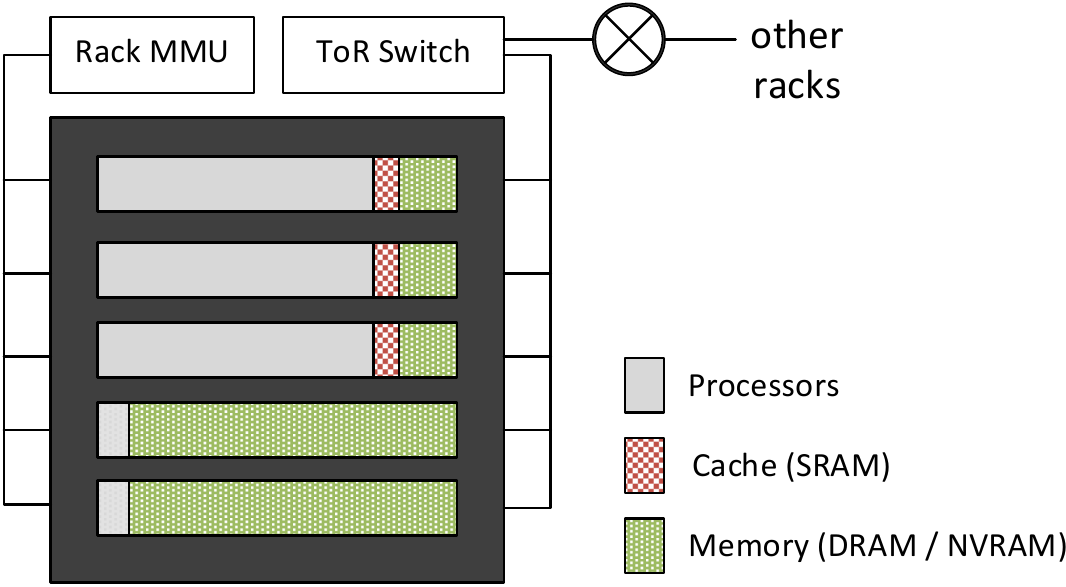}
\caption{Proposed architecture for disaggregated data centers (DDCs).
    Racks consist of blades housing compute or memory elements that
    are connected through the \rackmmu. Compute elements have caches
    and some local memory, while memory elements rely on
    an attached processor to mediate accesses. Network communication between
    compute elements within and across racks uses the Top-of-Rack Ethernet 
    switch.}%
\label{f:rack}
\end{figure}

In the absence of existing disaggregated data centers, a number of different 
  architectures have been proposed~\cite{themachine15,yizhou18legoos,
  lim09disaggregated,lim11disaggregated,novakovic14scale}. 
While these architectures differ in some of the details, the general
  strokes are similar.
We assume the architecture given in Figure~\ref{f:rack}, which
  has three core components: individual blades with \emph{compute
  elements} and \emph{memory elements}, connected over a low-latency
  \emph{programmable resource interconnect}.  
While we have chosen to explore these ideas in the context of a single 
  architecture for simplicity, we believe that they are broadly applicable 
  to other disaggregation models.

\paragraph{Compute elements.} 
The basic compute elements in our rack are commodity
  processors which retain the existing memory hierarchy with
  private core and shared socket caches.
While some architectures have processors operate entirely on remote 
  memory~\cite{novakovic14scale}, this requires major modifications to
  the processor to support instructions such as \texttt{PUSH} and
  \texttt{POP} that implicitly reference the stack, as well as to
  the memory and caching subsystems.
In line with the majority of proposed architectures, we assume a 
  small amount of locally-attached memory at the compute elements,
  which is used for the operating system and as a
  small cache to improve performance~\cite{gao16network, yizhou18legoos}.

\paragraph{Memory elements.} 
Memory elements, which are conventional DRAM or NVRAM chips, can 
  be exposed directly across the interconnect (Fabric-Attached Memory) or
  fronted by a low-power processing element (e.g., mobile 
  processor, FPGA, or ASIC) that interacts with memory through a 
  standard MMU (Proxied-Memory).
We assume a form of proxied-memory where addressing, virtualization, and
  access control are delegated to the local processing element which interposes
  on memory requests.
Similar functionality can be achieved for fabric-attached memory
  by coordinating memory controllers at multiple compute elements.

\paragraph{Resource interconnect.} 
The resource interconnect allows processor and memory elements to
  communicate and can be based on RDMA over InfiniBand or Ethernet,
  Omnipath~\cite{omnipath}, Gen-Z~\cite{genz}, or a switched PCIe
  fabric~\cite{liqid17, liqid18}.
Our design is agnostic to the physical layer, but we assume a degree of 
  programmability and on-the-fly reconfiguration within the
  interconnect (that we call the \rackmmu) that allows
  compute and memory elements to be dynamically connected and disconnected
  in arbitrary configurations.
Recent work~\cite{shrivastav19shoal} proposes and implements one such network
  fabric; although the proposed architecture lacks a programmable switch, it
  emulates its functionality through a Clos network of switches and a
  coordination-free scheduling protocol.

\paragraph{Resource partitioning and allocation.} 
We assume that the unit for disaggregation 
  is a single rack (i.e., compute and memory elements reside in the same 
  rack), with resources being partitioned into the desired compute
  abstractions, such as virtual machines, containers, or processes, and
  presented to applications (we generically refer to all of these
  compute abstractions, which host application workloads, as processes).
The \rackmmu{} acts as a resource manager for the rack and
  is responsible for resource partitioning within the rack
  and assigning compute and memory elements to processes.

The \rackmmu{} has a similar policy regarding sharing of hardware
  resources as LegoOS~\cite{yizhou18legoos}: processes may share
  the same memory element, but not the same regions of memory (i.e., there
  is no shared memory).
Similarly, compute elements can host multiple processes, but all the
  threads of a process are restricted to a single compute element.
This simplifies caching, as shared memory would require coherence
  across the local memory attached to the compute elements.
Memory is allocated at a fixed page-sized granularity, which is chosen
  according to the addressing architecture of compute and memory elements.
The \rackmmu{} is responsible for high-level placement decisions for
  processes and picks compute and memory elements on the basis of some
  bin-packing policy, while fine-grained sharing and isolation across
  co-hosted processes are managed by the local OS.

\paragraph{Addressing and access control.} 
Traditional processes expect to operate on a private virtual address
  space, regardless of the physical layout of the underlying memory. 
To preserve this illusion, the \rackmmu{} stores a virtual-to-physical
  (V2P) mapping for each process, which resembles a traditional
  per-process page table.  
Compute elements query this V2P mapping, which may be cached locally, 
  to route requests to the correct memory element.

The \rackmmu{} is also responsible for configuring access control to
  memory.
When memory is allocated, the \rackmmu{} ensures that the topology of
  the interconnect allows for the existence of a path between the
  corresponding compute and memory elements.
It also configures the page tables at the memory elements with
  the process identifier (effectively the \texttt{CR3}),
  the virtual address, and the appropriate permissions,
  enabling local enforcement at the memory elements.
While this mechanism is specific to proxied memory,
  fabric-attached memory systems have proposed a capability-based
  protection system to achieve similar 
  functionality~\cite{acherman17separating}.

\paragraph{Scaling out.} 
Not all applications want to live within a single rack: to span racks,
  traditional Ethernet-based networking is available through a commodity
  top-of-rack (ToR) switch that connects to the rest of the data
  center network. 
Distributed applications comprising multiple processes have to choose
  the appropriate deployment: intra-rack deployments enjoy
  lower latencies, while cross-rack deployments have greater failure 
  independence.
This decision is analogous to the one faced by developers when
  selecting the appropriate placement group~\cite{placementgroups} or
  availability set~\cite{availabilitysets} in cloud deployments today.

\section{Exposing disaggregation}\label{s:pl}

In traditional architectures, the OS is responsible for managing hardware 
  resources, allocating them to processes, and enforcing isolation of 
  shared resources.
In a disaggregated environment, this is no longer true and
  resource allocation is now within the bailiwick of the \rackmmu;
  the local OS at compute elements continues to be responsible for managing
  the underlying hardware, providing local scheduling and isolation,
  and presenting a standard programming interface to applications.
Additionally, the OS is responsible for transparently synchronizing
  application state between local and remote memory and, if any state
  is locally cached, managing the contents and coherence of this
  cache~\cite{infiniswap, yizhou18legoos}.

Prior OSes for DDCs~\cite{yizhou18legoos, carbonari17tolerating}
  have chosen to implement a standard POSIX API and abstract
  away the disaggregated nature of DDCs from applications.
While this allows existing unmodified applications to run on DDCs,
  our case studies~(\S\ref{s:parallel} and \S\ref{s:paxos}) argue
  that many of these applications could achieve better performance
  if they had more visibility and control.
Accordingly, we advocate for the design and implementation of
  the following three operations as OS interfaces.

\subsection{Memory grant}\label{s:grant}
Memory is reassigned at page granularity by moving it from
  the V2P mapping of one process to another at the resource manager (\rackmmu)
  and invalidating any cached V2P mappings at compute elements. 
Following this, the resource manager revokes access to that memory region by 
  modifying the page table entries for proxied memory (or by generating a 
  new capability for fabric-attached memory); the detached memory can 
  then be attached to an existing process similar to newly allocated memory.

Memory reassignment is conceptually similar to a memory \textit{grant}
  operation in L4~\cite{l4}, with one significant difference: as reassigned
  pages may contain data structures with internal references, these pages 
  must be attached to the same virtual address to prevent dangling pointers.
To avoid a situation in which the receiving process has already used the
  provided virtual addresses (which would create ambiguity), we 
  propose reserving a fixed number of bits of the virtual address to act as 
  a process identifier.

Mechanistically, we envision memory reassignment to occur, like in L4, 
  through message passing between OS instances.
This transfer is initiated by the application through a system call
  similar to \texttt{vmsplice()} in Linux: when called with
  the \texttt{SPLICE\_F\_GIFT} flag, the process ``gifts''
  the memory to the kernel, promising to never access it again.
As the page continues to use the same virtual address space in the
  receiving process, the sender OS marks the virtual address as
  being ``in use'' and prevents further allocations or mappings to it.
Receiving processes are notified about the addition of new pages by
  their local OS through signals.

\subsection{Memory steal}\label{s:steal}

Memory grants are the most natural flavor of memory reassignment, but are
  not particularly useful in the case of compute element failures.
An alternative is for other entities (processes or local OSes) to be able to 
  take away, or \emph{steal}, a process' memory.
For example, when a compute element crashes, another process belonging
  to the same application could request the crashed process' memory.
This is similar to how servers in Frangipani~\cite{thekkath97frangipani} keep 
  their logs remotely, and can request the logs of servers that have crashed 
  to resume their operations.

Two questions naturally arise in this case: firstly, who is allowed to
  trigger memory reassignments and when is it acceptable to do?
Secondly, how does the application guarantee the semantic consistency of
  memory that may abruptly be stolen?
While it is clear in the context of memory grants that a process should
  have the authority to give away its own memory, the policy around
  forcible reassignment is less clear.
One possibility is to group trusted processes together and allow any
  group member to initiate reassignment; another is to require
  a group of processes reach consensus before reassigning any memory.
In terms of timing, while we envision this as primarily an aid to recovery
  mechanisms when a process has crashed (or is suspected of having crashed),
  there might be applications where stealing memory from a running process
  is acceptable and actually profitable.

We propose to expose memory stealing via a syscall that requires the id
  of the source process and uses the group of the calling process as a
  capability for authentication; memory allocated using \texttt{brk} or
  \texttt{mmap} can disallow future reallocation with the appropriate
  flags.
We do not enforce a specific policy at the \rackmmu{} and instead
  leave it up to the application to determine what is appropriate (we
  explore one such policy in the context of Paxos in
  Section~\ref{s:dispaxos}).
While a buggy application can mistakenly steal its own memory and crash,
  this is not morally different from threads stomping on each
  other's memory in buggy shared memory applications.

The second challenge is maintaining crash consistency for reassignments.
This is non-trivial since most applications are not written with the idea
  that memory should be consistent and all invariants maintained at every
  point in the midst of computation; most applications do in fact have
  temporary windows of inconsistency.
While certain programming abstractions such as transactional
  memory and objects~\cite{shavit95software, herman16type}
  provide atomicity, they are not sufficient in the case
  of compute element failures.
Storage systems have historically faced similar challenges in allowing
  application state to outlive compute and building transactional,
  crash consistent programming models for non-volatile memory (NVRAM)
  is an active area of
  research~\cite{wu94envy, mnemosyne, nvheaps, cdds, xu16nova, kaminotx}.
Applications can adopt any of these programming models, which rely on
  a combination of techniques such as journaling, soft
  updates~\cite{softupdates}, shadow copies~\cite{bpfs}, and undo
  logs~\cite{nvheaps} to remain crash consistent when updating
  structures in remote memory.

Nevertheless, even when these structures are consistent in
  remote memory, the metadata required to locate them may be in
  processor registers, caches, or in stack variables that are not part
  of remote memory.
Applications typically do not have a namespace to locate internal objects
  and instead rely on the compiler to keep track of them; consequently,
  when memory is reassigned to a new process, finding
  the necessary objects from raw memory pages would be a momentous task,
  akin to searching for a lost treasure.

Our suggestions for this are two-fold: first, applications can use an
  asynchronous, event-based model that forces them to reason about
  \textit{all} critical state and package it into a heap object before
  yielding (i.e., stack ripping~\cite{stackrip}), since that is all that
  persists across invocations.
Secondly, the application can use a file system like namespace for
  objects~\cite{pmfs}, or it can distribute metadata about
  heap objects (depending on the application, this could be as minimal
  as the root address of a tree), in anticipation of failures, that act
  as a ``map'' to help locate critical state.
  
\subsection{Failure notification}\label{s:failure}

Compute elements should be notified about memory failures either
  asynchronously using liveness information from a reliable interconnect
  or explicitly in response to accesses on unreliable interconnects.
In the latter case, compute elements can receive messages from the
  controller of the memory element (when specific elements have
  failed), or rely on timeouts (when the entire memory element is
  unreachable).
Error notifications are propagated back to the application through
  OS signals (\texttt{SIGBUS}); applications that want to manage faults
  can register for these signals and trigger a failure-recovery protocol,
  while legacy applications may safely ignore them.

As memory failures may result in the loss of application state, it is
  sometimes unclear how an application should leverage failure
  notifications.
To guard against such cases, an application can pre-register a group of
  processes with the OS that will be informed in case of failures (these
  processes essentially serve as ``emergency contacts'').
This group is stored in a per-process forwarding table within the OS.
As the OS is local to the compute element, memory failures do
  \textit{not} affect the forwarding table; consequently, the
  application can defer notification to the OS using a syscall which
  broadcasts the error to the corresponding group.
This allows other processes to learn of the failure and respond
  appropriately, making the compute element a local failure
  informer~\cite{aguilera09no, leners11detecting}.

Failures of compute elements are harder to detect as the absence of
  accesses to a particular memory element need not be a sign of failure.
We propose the addition of a rack-level monitor that periodically
  verifies the health of compute elements using heartbeats and
  triggers the appropriate action when failures are detected.
Applications can register a group of processes to inform in the case
  of failures, similar to the groups registered for memory failures;
  alternatively, they can also register a lightweight failure handler
  to be run, in an isolated context, at the monitor.
While this monitor is a single point of failure and may not detect
  all failures, we view it as an optimization, rather than a replacement,
  to failure detection using end-to-end timeouts at the application.

\begin{figure}[t]
\centering
{\small
  \begin{tabular*}{\columnwidth}{@{\extracolsep{\fill}}l c} %l} 
    \textbf{setting} & \textbf{mean RTT ($\mu$s)}\\ 
    \toprule
    Cross-rack (Cloud) & 45 \\
    \midrule
    Intra-rack (eRPC~\cite{erpc}) & 2 \\
    \midrule
    Future intra-rack (Mellanox ConnectX-6~\cite{mellanox}) & 1 \\
    \bottomrule
  \end{tabular*}
}
\caption{Comparison of the latency of data transfer between VMs in the
  same and on different racks. The cross-rack number is derived
  experimentally and represents the mean round-trip time (RTT) between
  two VMs, with accelerated networking, within a cloud data center. We
  ensure that VMs are placed on different racks using the
  appropriate availability primitives~\cite{availabilitysets,
  placementgroups}. Both current and future intra-rack numbers are
  taken from the referenced publications.}
\label{f:micro}
\end{figure}

One might wonder why local failure informers are better than just
  using application-level timeouts to detect failures---especially
  given the reliance on timeouts to detect compute and memory failures.
The answer is that we can exploit the difference between intra-rack and 
  cross-rack latencies; as we show in Figure~\ref{f:micro}, this difference is 
  a few orders of magnitude.
As compute and memory are located within the same rack, we make the
  assumption that the \rackmmu{} achieves comparable latencies.
This allows local failure detectors to have more aggressive
  timeouts and trigger recovery procedures earlier.\footnote{The exact
  gains are hard to quantify since network latency is only one out of many
  factors considered when setting end-to-end timeouts~\cite{aguilera09no}.}

\subsection{Feasibility of implementing the \rackmmu}\label{s:challenges-rmmu}

The memory interconnect described so far is capable of routing requests
  between any compute and memory elements within the rack, as well as
  blocking communication between any such elements, at very low latency.
It has enough space to store address mappings for each process,
  so that accesses from compute elements are transparently routed to
  the correct memory element; further, it supports dynamic
  reconfiguration of routes and mappings without requiring any downtime.

While existing research and production hardware satisfies some of these
  requirements, achieving their composition remains an open problem.
Programmable switches, such as the Barefoot Tofino and Cavium XPliant,
  offer low-latency, reconfigurable routing between compute and memory
  elements, but are limited in their port counts and memory, restricting
  their scale.
In contrast, Shoal~\cite{shrivastav19shoal} supports high-density racks with 
  hundreds of compute and memory elements, but does not currently offer the 
  low latency, programmability, and reconfigurability required for 
  \texttt{grant} and \texttt{steal} operations.

\section{Case study: Paxos}\label{s:paxos}\label{s:paxos:bg}

Applications use Paxos~\cite{lamport98part} to tolerate failures via 
  the replicated state machine approach~\cite{lamport78time, oki88viewstamped, 
  schneider90implementing}: Paxos ensures that different replicas (which are 
  deterministic state machines that implement the application's logic) 
  execute the same commands in the same order, ensuring that all replicas 
  transition through the same sequence of states.
If a replica fails, a client can simply issue its requests to a live replica.

Replica failures lead the system into a state of \emph{reconfiguration}
  where the old failed replica is removed and 
  a new replica is introduced~\cite{lamport09vertical, lamport10reconfiguring,
  chandra07paxos}.
This prevents too many failures from accumulating over time and making the
  system unavailable.
Mechanistically, reconfiguration achieves two goals: first,
  it brings new replicas up to date by having them fetch the latest state
  from existing replicas or persistent storage~\cite{chandra07paxos}.
Second, it prevents old replicas that have been excluded from the current 
  configuration (presumably because they have failed) from participating if 
  they come back online.

\paragraph{Detecting failures.}\label{s:paxos:detection}

Detecting failures is a challenging proposition in an
  asynchronous environment due to the difficulty of distinguishing between 
  crashed and slow processes~\cite{flpimpossible, aguilera09no}.
Consequently, Paxos implementations rely on heartbeats
  and keep-alives with conservative end-to-end timeouts to ascertain the
  state of processes.
Recent failure detectors~\cite{leners11detecting, leners13improving, 
  leners15taming} quickly and reliably detect failures and kickstart 
  recovery mechanisms in asynchronous settings using a combination of local,
  host-based monitors that track the health of components across the
  stack, and lethal force.
In cases where failures are suspected but cannot be confirmed, these 
  detectors forcibly kill the process---the intuition
  behind this protocol (called \texttt{\small STONITH} or ``Shoot the Other
  Node in the Head'') is that unnecessary failures are preferable to 
  uncertainty.

\subsection{Paxos reconfiguration in DDCs}\label{s:dispaxos}

The failure independence of DDCs enables new ways to detect and
  recover from failures in fault-tolerant applications using Paxos.
We assume that the replicas of this application run in different racks
  within the same data center---a reasonable assumption for
  applications that want greater failure independence without paying
  the costs of wide-area traffic.
Within this deployment, we explore two scenarios: a compute element
  that loses some or all of its memory elements, and a faulty compute
  element with functional memory elements.

\paragraph{Dead compute with live memory.}\label{s:paxos:dead-comp}

When a replica dies, one could in principle reassign the state machine's
  memory to another compute element and the system could continue operating 
  unimpeded.
Such reassignment effectively \textit{reincarnates} the old node, from 
  the perspective of Paxos, which allows the consensus 
  group to return to full health faster (no need to retrieve the state from 
  a checkpoint or another replica). 
However, as we discuss in Section~\ref{s:steal}, the developer must
  ensure that the state machine's transition function preserves memory 
  consistency after crashes.

Should the failure of the compute element be detected faster
  than the end-to-end timeout of the Paxos group---a likely scenario
  due to the difference between intra- and cross-rack latencies---the 
  reincarnation can be transparent to the rest of the system.
In such cases, a client and other replicas will only observe a connection
  termination and will attempt to reconnect.
Each Paxos replica registers a \textit{fast failure handler} with the rack
  monitor that requests the \rackmmu{} to
  provision a new replica that can take ownership of the dead replica's
  memory with the \texttt{steal} operator of Section~\ref{s:steal}.
Failures detected by the monitor trigger this handler, while
  undetected failures are eventually detected by another replica which
  must gain consensus, either through a proposal or from a stable
  leader, before reincarnating the failed instance.

In response to a \texttt{steal} operation, the \rackmmu{} revokes and reassigns
  access to the region of memory.
Revocation is needed because compute element failures are not always
  fail-stop and the system must prevent a temporarily unavailable
  compute element from returning and corrupting state.
The ToR switch can redirect cross-rack traffic to the new compute element
  using OpenFlow rules; further, it can also use these rules to fence the old
  compute element off from the rest of the network~\cite{leners15taming}.

\paragraph{Dead memory with live compute.}\label{s:paxos:dead-mem}

When a compute element flushes its operations to a remote
  memory element, it is possible for this operation to fail if the
  memory element is down.
Instead of terminating the application right away, as we discuss in
  Section~\ref{s:failure}, the OS propagates a signal up the stack
  or forwards the signal to other replicas.
This mechanism allows other replicas to detect memory
  failures more quickly than relying on end-to-end timeouts.
Indeed, were the application to be terminated immediately without this
  notification, other replicas would not know whether the memory remains alive
  or not, leading to ambiguity as to the type of failure.

\section{Case study: Data parallel computations}
\label{s:parallel}

In-memory data parallel frameworks such as data flow and graph 
  processing systems~\cite{murray13naiad, zaharia12resilient, 
  dean04mapreduce, isard07dryad, gonzalez12powergraph, malewicz10pregel} 
  express computations as a series of nodes, where each 
  node performs an operation on its inputs.
In these systems, it is often necessary to move data between nodes so that the 
  output of a node may be used as the input to the next node.
For example, in MapReduce~\cite{dean04mapreduce}, the output of mappers
  is shuffled and sent to reducers that operate on a chunk of related data.

Applications represent large compute jobs as a set of smaller tasks,
  and distribute these tasks across nodes using the data parallel framework.
While completing these jobs requires all the individual tasks to
  finish, tasks are often unexpectedly delayed due to factors such as
  load imbalances and workload skews, failures, and hardware defects.
As such \emph{stragglers} hold up the entire job and significantly
  impact completion rates, frameworks employ a variety of mitigation
  techniques including blacklisting slow machines, speculatively timing
  out and rerunning tasks~\cite{dean04mapreduce, mantri}, and even
  proactively launching multiple replicas of the same task~\cite{dolly}.

We believe that executing data parallel systems transparently on a DDC
  would leave performance on the table, and argue for the use the operators
  described in Section~\ref{s:pl} to speed up data movement and
  straggler mitigation.

\begin{figure}
\centering
\includegraphics[width=0.6\columnwidth]{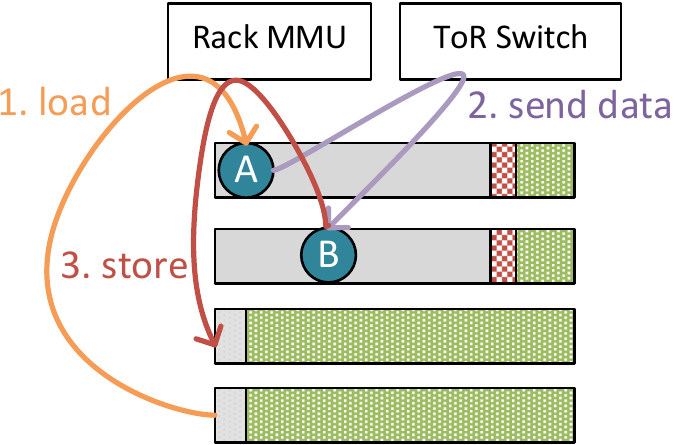}
\caption{Data transfer between two nodes, A and B, in the same rack. This
  figure assumes that not all of the data is cached in blade-local memory.
  (1) Node A loads its data from the remote memory via the \rackmmu. 
  (2) Node A sends the data to Node B over a TCP/IP stream via the ToR switch. 
  (3) Node B stores the data into its remote memory via the \rackmmu.}%
\label{f:parallel}
\end{figure}

\paragraph{Faster data movement.}
Deploying an unmodified data parallel framework on a transparent DDC
  results in unnecessary data movement between computational nodes; for
  example, Figure~\ref{f:parallel} shows how transferring data between 
  nodes forces 3 network and memory RTTs.
First, the source processor fetches data from its remote
  memory over the memory interconnect.
Then, the source processor sends this data over the network to the destination 
  processor via the ToR switch.
Finally, the destination processor forwards the data via the memory 
  interconnect to the remote memory for storage.

Meanwhile, data transfer is often a bottleneck in these systems.
As an example, McSherry and Schwarzkopf~\cite{mcsherry15impact}
  demonstrate that Timely Dataflow~\cite{timely} achieves up to
  3$\times$ higher throughput when provided with a faster network.
Memory grants convert the 3 RTTs for data transfer into a single RTT over
  the memory interconnect.
The source, $A$, would call \texttt{grant} on the memory pages storing
  the data that it plans to send to the destination, $B$, and would
  indicate $B$ as the recipient of these pages; the \rackmmu{} would
  make the necessary adjustments to page permissions before notifying
  $B$ that the pages are ready to be mapped into its local address
  space.
Here, all the data transfer consists of small control messages that occur via 
  the \rackmmu{} and bypasses the slower ToR.

\paragraph{Dealing with stragglers.}

Straggler nodes in data parallel systems can have their memory
  forcibly reassigned to another node by having the job orchestrator
  \texttt{steal} the appropriate memory pages.
The recipient node can resume and complete the half-completed
  computation, rather than starting from scratch.
In case of failures, as with the Paxos example~(\S\ref{s:paxos:dead-mem}),
  the failure notification interfaces can inform the job orchestrator,
  allowing it to relaunch the task more quickly than relying on
  an end-to-end timeout.
If only the compute elements of the node have failed, the newly
  launched task can resume computation from where it had stalled.

\section{Discussion}\label{s:disc}

We are by no means the first to observe either the ability to reassign
  memory across processes or the failure independence of resources in
  DDCs.
While recent works on disaggregated systems have advocated for
  transparent solutions---RAID-style~\cite{patterson88case} memory
  replication in LegoOS~\cite{yizhou18legoos} and replication and
  switch-based failover by \cnb~\cite{carbonari17tolerating}---this has
  largely been driven by the desire to benefit legacy applications.
\cnb{} also observe that applications could benefit from information
  about failures but do not go further; we build on that observation
  and look at how applications that eschew transparency could use this
  information.
More specifically, we borrow ideas from systems for single-host
  IPC~\cite{lrpc, urpc, liedtke93, liedtke95, l4}, distributed shared
  memory~\cite{hudaklidsm, treadmarks, opal, cashmere}, and accelerated
  RPCs~\cite{fasst, erpc} for fast, zero-copy data transfer and from
  reliable failure informers~\cite{aguilera09no, leners11detecting,
  leners13improving, leners15taming} for faster recovery.

Disaggregation represents a fundamental change in how hardware
  resources are built, provisioned, and presented to applications for
  consumption.
As befitting an operator-driven initiative, early research has focused
  on changes necessary within the hardware %and interconnect,
  rather than the application.
But application developers are not averse to major changes in their
  programming model as long as they receive commensurate benefits; in
  fact, as witnessed by the prevalence of MapReduce, good %programming
  models can help applications transition more smoothly.
Reasoning about memory \texttt{grant}s and \texttt{steal}s is a significant 
  departure from existing programming models, but there is encouraging
  precedent: the Rust programming language successfully introduced ownership 
  and move semantics to guarantee memory safety and data race freedom.
We believe that similar abstractions could be useful in our context.

\subsection*{Acknowledgements}
We thank Andrew Baumann, Natacha Crooks, Joshua Leners, Youngjin Kwon, Srinath
  Setty, and Nathan Taylor for feedback and discussions that improved this 
  paper.

\frenchspacing

\begin{small}
\begin{flushleft}
\setlength{\parskip}{0pt}
\setlength{\itemsep}{0pt}
\bibliographystyle{abbrv}
\bibliography{conferences,paper} 
\end{flushleft}
\end{small}
\end{document}